\documentclass[12pt, a4paper]{article}
\usepackage[font=small,format=plain,labelfont=bf,up,textfont=normal,up,justification=justified,singlelinecheck=false]{caption}
\usepackage{subcaption}
\usepackage{mathrsfs}
\usepackage[a4paper, left=2cm,right=2cm]{geometry}
\usepackage[colorlinks=true,linkcolor=black,citecolor=teal,urlcolor=MidnightBlue,filecolor=black]{hyperref}
\usepackage{amsfonts}
\usepackage{amsmath,amssymb}
\usepackage{setspace}
\usepackage{slashed}
\usepackage{braket}
\usepackage[dvipsnames]{xcolor}
\definecolor{SchoolColor}{rgb}{0.6471, 0.1098, 0.1882}
\usepackage{subfloat}
\usepackage{tensor}
\usepackage{cite}
\usepackage{tikz}
\usepackage[thicklines]{cancel}
\usetikzlibrary{calc}
\usetikzlibrary{patterns}
\usetikzlibrary{arrows.meta}
\usepackage{graphicx}
\usepackage{bm}
\allowdisplaybreaks[4]
\usepackage{tensor}
\usepackage{cite}
\usepackage{tikz}
\usepackage{graphicx}
\graphicspath{{figure/}}
\usepackage{array}
\usepackage{booktabs}
\usepackage{multirow}
\usepackage{dcolumn}
\usepackage{bm}
\usepackage{verbatim}
\usepackage{textcomp}
\usepackage{graphicx} 

\numberwithin{equation}{section}
\newcommand{\bea}{\begin{eqnarray}}
\newcommand{\eea}{\end{eqnarray}}
\newcommand{\be}{\begin{equation}}
\newcommand{\ee}{\end{equation}}
\newcommand{\bs}{\begin{subequations}}
\newcommand{\es}{\end{subequations}}
\def\nn{\nonumber}

\newcommand{\beqs}{\begin{eqnarray}}
\newcommand{\eeqs}{\end{eqnarray}}
\numberwithin{equation}{section}

\newcommand{\Rmnum}[1]{\uppercase\expandafter{\romannumeral #1\relax}}

\def\c.c.{\mathrm{c.c.}}

\def\ep{\epsilon}

\def\om{\omega}
\def\Om{\Omega}

\newcommand{\zxh}{\color{red}}
\begin{document}
\begin{titlepage}

\begin{flushright}\vspace{-3cm}
{\small\today }
\end{flushright}
\vspace{0.5cm}
\begin{center}
	{{ \LARGE{\bf Reduction of topological invariants \vspace{8pt}\\ on null hypersurfaces}}}\vspace{5mm}
	
	\centerline{ Jiang Long\footnote{longjiang@hust.edu.cn} \quad  \&\quad Xin-Hao Zhou\footnote{d202580129@hust.edu.cn}}
	\vspace{2mm}
	\normalsize
	\bigskip\medskip
	\textit{School of Physics, Huazhong University of Science and Technology, \\ Luoyu Road 1037, Wuhan, Hubei 430074, China
	}
	\vspace{25mm}
	
	\begin{abstract}
		\noindent
		{Gravitational helicity flux density  represents  the angular distribution of helicity flux in general relativity. In this work, we explore its relationship to the reduction of topological invariants at future null infinity. Contrary to initial expectations, the Pontryagin term, which contributes to the gravitational chiral anomaly, is not related to the gravitational helicity flux density. Instead, the Nieh-Yan term, another topological invariant within the teleparallel equivalent of general relativity (TEGR), can reproduce this flux density. We also reduce these two topological invariants to null hypersurfaces describing near-horizon geometry. In this near-horizon context, the Pontryagin term yields a non-trivial quantity that may characterize a  Carrollian fluid helicity relevant to near-horizon physics. }
	\end{abstract}
\end{center}

\end{titlepage}
\tableofcontents
\section{Introduction}
Recently, a topological invariant, which is known as Chern number, has been reduced to the null infinity of asymptotically flat spacetime within electromagnetic theory \cite{Liu:2024rvz}.  Interestingly, under specific fall-off conditions, the boundary term at future/past null infinity corresponds precisely to the radiative helicity, from which one can extract the helicity flux density characterizing the angular dependence of the radiative helicity flux.  The result is novel in the following aspects. First, the helicity flux density can be obtained from the  superrotation operator commutator at future null infinity for spinning fields \cite{Liu:2023qtr,Liu:2023gwa,Liu:2023jnc}. Moreover, for fermionic theory, a similar  helicity flux  operator arises from the non-closure of the Lie transport of a spinor around a loop \cite{Guo:2024qzv}. Third, these helicity flux  operators generate nonlinear spin memory effects in both gravitational and electromagnetic theories\cite{Seraj:2022qyt,oblak2024orientation,Maleknejad:2023nyh}. Fourth, the second Chern class directly contributes to the ABJ anomaly, whose significance is established by the Atiyah-Singer index theorem \cite{Atiyah:1975jf}. Finally, while the Chern number typically enumerates  instanton numbers in Euclidean spacetime, the helicity flux operator, originating from the reduction of Chern number to null infinity within electromagnetic theory,  encodes dynamical information of the radiative fields.
   Consequently, extending the bulk reduction methodology \cite{Liu:2022mne,Liu:2024nkc} to reduce diverse topological invariants to null infinity and general null hypersurfaces is important. This approach promises analogous helicity flux operators in gravitational theory and potential discovery of new observables.

Topological invariants constitute significant entities across mathematics and physics. In four-dimensional gravity, classical invariants include the Pontryagin and Euler terms \cite{diffeo}, both fundamental to manifold topology.
\iffalse 
Topological invariants are interesting quantities in mathematics and
physics from both directions. 
Classical topological invariants in four-dimensional gravity include the so-called Pontryagin  and Euler characteristics\cite{diffeo}, both of which are fundamental objects in the characterization of the topological structure of a manifold \footnote{In Euclidean space, one can ultilize the metric to define the open ball and then it induces the usual topology. However, the Minkowski metric has the signature $(1,3)$ and thus the definition of the topology for Lorentz manifold becomes messy \cite{Zeeman1967TheTO,1976JMP....17..174H}. We will adapt a practical point of view and just use the same form of the topological invariants as in Euclidean space. }. \fi More precisely, considering four-dimensional spacetime $M$ and denoting the curvature 2-form as $R^{ab}$ and then the Pontryagin term $\tau$ and the Euler term $\chi$ are 
\bea 
\tau=\int_M R^{ab}\wedge R_{ab},\quad \chi=\int_M \epsilon_{abcd}R^{ab}\wedge R^{cd}
\eea respectively. We have ignored the normalization constant before the integrals. The Euler term, an alternating sum of Betti numbers, corresponds to the Gauss-Bonnet term in physics. One can find recent developments in \cite{Fernandes:2022zrq} on its role in gravitational theories. On the other hand, the first Pontryagin class induces chiral symmetry breaking of Weyl fermions via axial current anomalies \cite{PhysRevLett.37.1251}, analogous to ABJ anomalies in gauge theories \cite{PhysRev.177.2426,1969NCimA..60...47B}. 

This paper reduces key gravitational topological terms \cite{Kimura:1969iwz,Delbourgo:1972xb,Alvarez-Gaume:1983ihn} to null hypersurfaces. Table \ref{anomaly} summarizes three topological terms linked to chiral symmetry breaking.
\begin{table}
\begin{center}
\renewcommand\arraystretch{1.5}
 \begin{tabular}{|c||c|}\hline
Theory &Topological term\\\hline\hline
Gauge theory&$\int_M \text{tr}F\wedge F$\\\hline
Gravitational theory without torsion&$\int_M R^{ab}\wedge R_{ab}$\\\hline
Gravitational theory with torsion&$\int_M R^{ab}\wedge R_{ab},\quad \Lambda^2\int_M (T^a\wedge T_a-\theta_a\wedge \theta_b\wedge R^{ab})$\\\hline
\end{tabular}
\caption{\centering{Several topological terms.}}\label{anomaly}
\end{center}
\end{table} In gauge theories, the topological term is the 
Chern number and we have denoted the field strength 2-form as $F$. For torsion free gravity, the relevant topological term is the Pontryagin term. A longstanding debate concerns whether the Nieh-Yan term \cite{1982JMP....23..373N,1982AnPhy.138..237N,Nieh:2007zz} 
\be 
\zeta=\Lambda^2\int_M (T^a\wedge T_a-\theta_a\wedge \theta_b\wedge R^{ab})\label{NiehYan}
\ee
contributes to chiral anomalies in torsional gravity  \cite{PhysRevD.55.7580,PhysRevD.63.048501,Obukhov:1982da,Obukhov1997OnTC,Rasulian:2023gtb,Erdmenger:2024zty} where $\Lambda$ is a regularization cutoff and $T^a$ is the torsion 2-form and $\theta_a$ is the vielbein 1-form\footnote{In this paper, we will abstain from this debate. Our aim is rather to delineate several interesting features pertaining to the reduction of topological terms.}. Surprisingly, the reduction of the Pontryagin term to future null infinity yields a vanishing result.  In contrast, the radiative gravitational helicity in general relativity is recovered only through the reduction of the Nieh-Yan term in the teleparallel equivalent of general relativity (TEGR). Furthermore, a remarkable reversal of their roles occurs near event horizons. In this region, the  reduction of the Nieh-Yan term vanishes, while the Pontryagin term gives rise to a parity-odd quantity that could potentially characterize the helicity of a Carrollian fluid in near-horizon geometries.
 
The structure of this paper is as follows. In Section \ref{reduction}, we present the reduction of both the Pontryagin and Nieh-Yan terms to future null infinity. Our analysis focuses on the reduction method for the Nieh-Yan term in TEGR, establishing its connection to the gravitational helicity flux. Section \ref{extension} extends this framework to near-horizon null hypersurfaces and explores its connections to Carrollian fluid helicity. Finally, conclusions and future research directions are presented in the last section.

\section{Reduction of topological invariants to future null infinity}\label{reduction}
This section reduces the Pontryagin and Nieh-Yan terms to future null infinity.
\subsection{Pontryagin term}
 We will use Bondi coordinates $(u,r,x^A)$ near future null infinity where $u$ is the retarded time and $r$ denotes the radius. The angular coordinates are collected as $x^A=(\theta,\phi)$. The metric of an asymptotically flat spacetime may be denoted as 
\be 
ds^2=V du^2-2 B du dr+g_{AB}(dx^A-U^A du)(dx^B-U^B du).\label{metric}
\ee By imposing the standard fall-off conditions \cite{Bondi:1962px}, the undetermined functions can be solved near future null infinity order by order. The first few orders of the solution are
\begin{align}
    V&=-1+\frac{2M}{r}+\mathcal{O}(r^{-2}),\quad B=1+\mathcal{O}(r^{-2}),\quad U^A=-\frac{\nabla_B C^{AB}}{2r^2}+\mathcal{O}(r^{-3}),\\ g_{AB}&=r^2\gamma_{AB}+r C_{AB}+ \frac{1}{4}\gamma_{AB}C_{CD}C^{CD}+\mathcal{O}(r^{-1}).
\end{align} The shear tensor $C_{AB}$ is a symmetric, traceless and rank-two tensor whose indices are raised by the inverse of the metric $\gamma_{AB}$  of the unit sphere. The covariant derivative $\nabla_A$ is also adapted to the metric $\gamma_{AB}$. For simplicity, we do not include the conformal factor that deforms the unit sphere, as has been discussed in \cite{Barnich:2010eb}. In general, the Pontryagin term is a topological invariant for a non-trivial manifold and the first Pontryagin class 
\be 
p=R^{ab}\wedge R_{ab}
\ee is not an exact form globally. However, it is always possible to write it as an exterior derivative of a Chern-Simons 3-form locally. More precisely, in a local coordinate chart, we have
\be 
p=d\left(\omega^{ab}\wedge d\omega_{ab}-\frac{2}{3}\omega^a_{\ b}\wedge \omega^b_{\ c}\wedge \omega^c_{\ a}\right)
\ee where $\omega^{ab}$ is the spin connection of the manifold whose flat indices are raised and lowered by the flat metric $\eta_{ab}=\text{diag}(-1,+1,+1,+1)$ {with $a=0,1,\hat{A}$}. Using the Stokes' theorem, the Pontryagin term can be reduced to boundary terms 
\be 
\tau=\int_M R^{ab}\wedge R_{ab}=\int_{\partial M}\left(\omega^{ab}\wedge d\omega_{ab}-\frac{2}{3}\omega^a_{\ b}\wedge \omega^b_{\ c}\wedge \omega^c_{\ a}\right).
\ee 
Note that the boundary $\partial M$ consists of several parts. We focus here on the term at future null infinity
\be 
\text{CS}_+=\int_{\mathcal {I}^+} \left(\omega^{ab}\wedge d\omega_{ab}-\frac{2}{3}\omega^a_{\ b}\wedge \omega^b_{\ c}\wedge \omega^c_{\ a}\right).
\ee This is the bulk reduction of Pontryagin term to future null infinity. To be more precise, one may choose a constant $r$ slice $\mathcal H_r$ and then take the limit $r\to\infty$  while keeping the retarded time finite
\be 
\text{CS}_+=\lim_{r\to\infty,\ u \ \text{finite}}\int_{\mathcal H_r}
 \left(\omega^{ab}\wedge d\omega_{ab}-\frac{2}{3}\omega^a_{\ b}\wedge \omega^b_{\ c}\wedge \omega^c_{\ a}\right).\label{chernsimons}
\ee The evaluation of the Chern-Simons term is straightforward. One may choose the vielbein fields 
\begin{align}
    \theta^0&=\frac{1}{2}(B-V)du+dr,\quad \theta^1=\frac{1}{2}(B+V)du-dr,\quad \theta^{\hat A}=E_A^{\hat A}(dx^A-U^{ A}du)\label{vel}
\end{align} with $E_A^{\hat A}$ satisfying the following equations
\be 
E_A^{\hat A}E_B^{\hat B}\delta_{\hat A\hat B}=g_{AB},\quad E_A^{\hat A}E_B^{\hat B}g^{AB}=\delta^{\hat A\hat B}.
\ee One of the choices  of $E_A^{\hat A}$ would be 
\be 
E_A^{\hat A}=r\overline E_A^{\hat A}+\frac{1}{2}C_{A}^{\ \hat A}+\mathcal{O}(r^{-1})
\ee where $\overline E_A^{\hat A}$ is the vielbein associated with the metric of the unit sphere and $C_{A}^{\ \hat A}=\overline E^{B\hat A}C_{AB}$. By solving the torsion free equation 
\be
T^a=d\theta^a+\omega^a_{\ b}\wedge\theta^b=0,
\ee we find the fall-off behavior  of the spin connection 
\begin{align}
    \omega_{u\hspace{3pt}b}^{\ a}=\mathcal{O}(r^{-2}),\quad \omega_{r\hspace{3pt}b}^{\ a}=\mathcal{O}(r^{-2}),\quad \omega_{A\hspace{3pt} b}^{\ a}=\left(\begin{array}{ccc} 0&\mathcal{O}(r^{-1})&\mathcal{O}(1)\\ \mathcal{O}(r^{-1})&0&\mathcal{O}(1)\\ \mathcal{O}(1)&\mathcal{O}(1)&\mathcal{O}(1)\end{array}\right).
\end{align} The components that may  contribute to the boundary Chern-Simons term are computed as follows\begin{align}
    \omega_{A}^{\ 0\hat B}&=\dot C_A^{\ \hat B}+\mathcal{O}(r^{-1}),\quad \omega_A^{\ 1\hat B}=-\dot C_A^{\ \hat B}+\overline E_A^{\ \hat B}+\mathcal{O}(r^{-1}),\quad \omega_A^{\ \hat B\hat C}=\overline\omega_A^{\ \hat B\hat C}+\mathcal{O}(r^{-1}),
\end{align}
where $\dot C_A^{\ \hat B}=\partial_u C_A^{\ \hat B}$ and 
$\overline\omega_A^{\ \hat B\hat C}$ is the spin connection of the unit sphere. Substituting  the above into the equation \eqref{chernsimons}, we find a null result at future null infinity
\be 
\text{CS}_+=0.
\ee 
This null result arises from the cancellation of two finite terms
\be 
\int du\wedge dx^A\wedge dx^B (\omega_{A\hspace{3pt}\hat{ B}}^{\ 0}\wedge \dot{\omega}_{B\hspace{3pt}0}^{\ \hat B}+\omega_{A\hspace{3pt}\hat{ B}}^{\ 1}\wedge \dot{\omega}_{B\hspace{3pt}1}^{\ \hat B})=0.
\ee As a consistency check, we have also considered an equivalent form of the boundary Chern-Simons term 
\be 
\text{CS}_+=\lim_{r\to\infty,\ u\ \text{finite}}\int_{\mathcal H_r} d^3y \sqrt{h} n_\mu \epsilon^{\mu\nu\rho\sigma}\left(\Gamma^\alpha_{\nu\beta}\partial_\rho\Gamma^\beta_{\sigma\alpha}+\frac{2}{3}\Gamma^\alpha_{\nu\beta}\Gamma^\beta_{\rho\gamma}\Gamma^\gamma_{\sigma\alpha}\right)\label{csnull}
\ee where $\Gamma^\alpha_{\mu\beta}$ is the Christoffel connection. The coordinates of $\mathcal H_r$ are denoted by $y^i=(u,x^A)$ and $n_\mu$ is the normal covector and $h_{\mu\nu}$ is the induced metric of $\mathcal H_r$. As expected, the result is still zero in this formalism. 
\subsection{Nieh-Yan term}
We have demonstrated that the Pontryagin term yields a vanishing result at the boundary for asymptotically flat spacetimes. The challenge, therefore, lies in identifying other topological invariants that could give rise to a gravitational helicity flux operator, analogous to the Chern number in gauge theories. The Euler term can be ruled out, as it also fails to produce a non-trivial result at null infinity, consistent with the established claim that the Euler term is unrelated to chiral anomalies \cite{PhysRevLett.37.1251}. It thus appears that no classical topological invariants are associated with radiative gravitational helicity.

However, Cartan developed an extension of general relativity, now known as Einstein–Cartan (EC) theory \cite{Cartan:1923zea,Cartan:1924yea}, in which spacetime is permitted to possess both curvature and torsion. In EC theory, curvature is sourced by the stress energy tensor, while the non-dynamical torsion is sourced by the intrinsic spin density. Within this framework, general relativity is recovered as the limiting case with vanishing torsion. The role of torsion in gravity and its potential physical implications are discussed extensively in \cite{Cartan1979ElieCA,Kibble:1961ba,RevModPhys.48.393,2002PhR...357..113S}. 

We notice that Nieh-Yan term \eqref{NiehYan} is a topological invariant when the torsion is non-zero. Indeed, it is  locally exact 
\be 
\zeta=\Lambda^2\int_M d(\theta^a\wedge T_a).
\ee Therefore, by reducing this to the future null infinity, we obtain the quantity 
\be 
H=\Lambda^2 \int_{\mathcal I^+} \theta^a\wedge T_a.\label{H}
\ee Unfortunately, this boundary term still vanishes in general relativity due to the absence of torsion.

Nevertheless, we observe that the formulation of general relativity is not unique. An equivalent theory, known as the Teleparallel Equivalent of General Relativity (TEGR)\footnote{Meanwhile, there is another equivalent theory that uses nonmetricity  to describe gravitation with vanishing torsion and curvature, named symmetric teleparallel equivalent of
 general relativity (STEGR)\cite{Nester:1998mp}. TEGR, STEGR and GR together describe the same gravitational theory  by torsion, nonmetricity and curvature, respectively. This equivalence is called the geometric trinity of gravity \cite{BeltranJimenez:2019esp}.}, models the gravitational field not as spacetime curvature, but as a dynamical force.  TEGR originated in Einstein’s work on unified field theory \cite{Unzicker:2005in} and was fully developed in the 1970s \cite{Moller:1961,Pellegrini:1962,Hayashi:1979qx,PhysRevD.14.2521}.

In general relativity, gravitation is manifested as curvature, and torsion is identically zero. In contrast, TEGR attributes gravitation to torsion, while curvature is set to zero. Thus, gravitational interactions can be interpreted either as purely geometric (as in general relativity) or as a force mediated by torsion (as in TEGR). Although TEGR lacks the geometric interpretation of general relativity, it is dynamically equivalent and is founded on the translation gauge group, making its structure more analogous to that of conventional gauge theories. For further details, we refer readers to \cite{2013tgif.book.....A}. 

Even though the Nieh–Yan term is absent in general relativity, it can arise in TEGR. We therefore turn to TEGR and evaluate the term \eqref{H} at future null infinity. In this theory, torsion is non-vanishing 
\be 
T^a=d\theta^a+\Omega^a_{\ b}\wedge\theta^b,
\ee where the $\Omega^a_{\ b}$ is the spin connection. Here, we use an overline to denote the spin connection, which must satisfy the zero-curvature condition 
\be 
d \Omega^{ab}+ \Omega^a_{\ c}\wedge\Omega^{cb}=0,
\ee implying that the spin connection is pure gauge. However, the choice of spin connection in TEGR is subtle \cite{Kr_k_2019}. We will revisit this point later. Substituting the vielbeins \eqref{vel} into \eqref{H}, we find
\begin{align}
H=&\Lambda^2\int du\wedge dx^A\wedge dx^B\left(-\frac{1}{4}C_{AC}\dot C_B^{\ C}-\frac{1}{2}\nabla_A\nabla_CC^C_{\ B}-\bar U_{\hat C}(\partial_A\overline E_B^{\hat C}-\partial_B\overline E_A^{\hat C})\right)\nn\\&+\Lambda^2\int_{\mathcal {I}^+}\theta^a\wedge\Omega_{ab}\wedge\theta^b\label{Heq}
\end{align}
where $\bar U_{\hat A}=-\frac{1}{2}\nabla^B C_{B\hat A}$. The choice of ${\Omega}^{ab}$ is crucial. In pure-tetrad teleparallel gravity \cite{Maluf:2013gaa}, the spin connection is set to zero in all reference frames. Since the vielbein encodes both gravitational and inertial effects (the latter due to frame choice), it becomes necessary to eliminate purely inertial contributions, especially in Minkowski spacetime, a process sometimes referred to as regularization in teleparallel gravity \cite{PhysRevD.73.124017,Lucas:2009nq,Krssak:2015rqa}. To be more precise, near future null infinity, the metric is approximated by 
\be 
ds^2=-du^2-2du dr+r^2(d\theta^2+\sin^2\theta d\phi^2)
\ee which corresponds to $V=-1,\ B=1,\ U^A=0,\ g_{AB}=r^2\gamma_{AB}$ in \eqref{metric}. A preferred spin connection can be fixed by solving the torsion-free condition in flat space. The reference vielbeins analogous to \eqref{vel} are
\be 
\theta^0=du+dr,\quad \theta^1=-dr,\quad \theta^{\hat A}=r\overline{E}_A^{\ \hat A}dx^A
\ee and the associated spin connection is obtained by solving the torsion free equation 
\be 
\Omega_A^{\ \hat B\hat C}={r}\overline E_{A\hat A} \Omega^{\hat A\hat B\hat C},\quad \Omega^{\hat A\hat B\hat C}=\frac{1}{2}(\overline \alpha^{\hat B\hat C\hat A}+\overline\alpha^{\hat C\hat A\hat B}-\overline\alpha^{\hat A\hat B\hat C})
\ee where 
\be 
\overline\alpha^{\hat A\hat B\hat C}={ \frac{1}{r}}\overline E^{A\hat B}\overline E^{B\hat C}(\partial_A\overline E_B^{\ \hat A}-\partial_B\overline E_A^{\ \hat A}).
\ee With this spin connection, the equation \eqref{Heq} becomes
\begin{align} 
H=&\Lambda^2\int du \wedge dx^A\wedge dx^B\left(-\frac{1}{4}C_{AC}\dot C_B^{\ C}-\frac{1}{2}\nabla_A\nabla_CC^C_{\ B}\right)\nn\\=&-\frac{1}{4}\Lambda^2\int du d\Omega \epsilon^{AB} \left(C_{AC}\dot C_B^{\ C}-2\nabla_A\nabla_C C^C_{\ B}\right).
\end{align}
The integrand corresponds precisely to the spin precession rate of a gyroscope induced by gravitational waves \cite{Seraj:2022qyt}, including both linear and nonlinear terms in the shear tensor. After integration by parts, only the quadratic term remains, and the integrand reduces to the gravitational helicity flux density \cite{Liu:2023jnc, Long:2024yvj}
\be 
O(u,\Omega)=\frac{1}{32\pi G}\epsilon^{AB}C_{AC}\dot C_B^{\ C}
\ee with the identification 
\be 
\Lambda^2=\frac{1}{8\pi G}.
\ee 

%How to define the spin connection \cite{PhysRevD.73.124017,Lucas:2009nq,Krssak:2015rqa}

\section{Extension to horizon}\label{extension}
In this section, we reduce the Pontryagin and Nieh-Yan terms to general null hypersurfaces. A typical example is the black hole event horizon, which can be effectively described using a membrane paradigm \cite{1982MNRAS.198..345M,1986bhmp.book.....T,1986PhRvD..33..915P} as a two-dimensional surface evolving in three-dimensional space. As argued in \cite{Penna:2015gza,Donnay:2019jiz}, near-horizon black hole physics may be understood as Carrollian hydrodynamics. This framework provides a macroscopic description based on Carrollian symmetry, contrasting with both relativistic and Galilean hydrodynamic formulations \cite{1959flme.book.....L,Kovtun:2012rj,Jeon:2015dfa,Romatschke:2017ejr}. For recent developments in Carrollian hydrodynamics, see \cite{Ciambelli:2018wre,Ciambelli:2018xat,Ciambelli:2018ojf,Petkou:2022bmz,Armas:2023dcz,Freidel:2022vjq,Redondo-Yuste:2022czg,Freidel:2022bai,Kolekar:2024cfg,Bagchi:2023ysc,Bagchi:2023rwd,Arenas-Henriquez:2025rpt}. Further aspects of black holes and Carrollian symmetry are discussed in \cite{Penna:2018gfx,Bagchi:2022iqb,Ecker:2023uwm,Bagchi:2023cfp,Chen:2024how,Tadros:2024bev,Aggarwal:2024yxy,Husnugil:2025edm}. Readers may also consult the review \cite{Bagchi:2025vri} for a comprehensive reference list. By exploring the reduction of topological invariants on null hypersurfaces, we may thus obtain interesting observables in the Carrollian fluid.

A suitable coordinate system for describing the near-horizon geometry is the null Gaussian coordinate system $(v,\rho,x^A)$, in which the metric takes the form\cite{1983CMaPh..89..387M} 
\begin{align}
    ds^2=-F dv^2+2dvd\rho+g_{AB}(dx^A+U^Adv)(dx^B+U^Bdv).
\end{align}
In component form, they are given by 
\begin{align}
    g_{vv}=-F+g_{AB}U^AU^B,\quad g_{v\rho}=1,\quad g_{vA}=U_A,\quad g_{AB}.
\end{align}
The components of the inverse metric are
\begin{align}
    g^{vv}=0,\quad g^{v\rho}=1,\quad g^{vA}=0,\quad g^{\rho\rho}=F,\quad g^{\rho A}=-U^A,\quad g^{AB}.
\end{align} The horizon $\overline M$ is located at $\rho=0$ and the near-horizon expansion of the metric is 
\begin{align}
    &F=2\kappa \rho+F_2\rho^2+\mathcal{O}(\rho^3),\quad U^A=\overline U^A\rho +{}^{(2)}\overline U^A\rho^2 +\mathcal{O}(\rho^3),\\ &g_{AB}=\gamma_{AB}+\lambda_{AB}\rho+ k_{AB}\rho^2+\mathcal{O}(\rho^3)
\end{align} where $\kappa=\kappa(v,x)$ is the surface gravity and $\overline U^A=\overline U^A(v,x)$ is the twist field. It is often convenient to take the hypersurface with fixed 
$\rho$, where the induced metric is 
\begin{align}
    ds^2|_{\rho =\text{const.}}=-Fdv^2+g_{AB}(dx^A+U^Adv)(dx^B+U^Bdv).
\end{align}
Taking the limit $\rho\rightarrow0$, thus the hypersurface becomes  horizon $\overline M$ and the metric turns into 
\begin{align}
    ds^2|_{\rho\rightarrow0}=-0\cdot \kappa dv^2+0\cdot \gamma_{AB}\overline U^Advdx^B+\gamma_{AB}(v,x)dx^A dx^B,
\end{align} which is degenerate with respect to the three-dimensional null hypersurface $\overline M$. Note that $\gamma_{AB}$ is also the metric of
spatial cross-section of the horizon and it can be used to raise and lower the indices of $(\overline U^A, \lambda_{AB}, k_{AB})$. In four dimensions, $\gamma_{AB}$ is two-dimensional, and one may choose the conformal gauge
\be 
\gamma_{AB}=e^{2\varphi(v,x)}\delta_{AB}.\label{gaugegamma}
\ee Note that $\varphi$ may depend on the time coordinate $v$. The second fundamental form on the event horizon is given by
\be 
K_{AB}=\frac{1}{2}\partial_v\gamma_{AB}=\frac{1}{2}\dot\gamma_{AB}=\dot\varphi\gamma_{AB},
\ee where $\dot{\varphi}=\partial_v \varphi$. Consequently, the expansion of the event horizon, which measures the dynamical variation of the area form associated with the spatial metric, is
\be 
\Theta=\gamma^{AB}K_{AB}=2\dot\varphi.
\ee Note that the shear tensor in conformal gauge is zero 
\be 
\sigma_{AB}=K_{AB}-\frac{\Theta}{2}\gamma_{AB}=0.
\ee Throughout this work, we do not employ the symmetric rank-2 tensors $\lambda_{AB}$ and $k_{AB}$. 

It is straightforward to compute the non-vanishing Christoffel symbols 
\begin{align}
    \Gamma^v_{vv}&=\frac{1}{2}\partial_\rho F-\frac{1}{2}\partial_\rho \left(g_{AB}U^AU^B\right),\\
   % \Gamma^v_{v\rho}&=0,\\
    \Gamma^v_{vA}&=-\frac{1}{2}\partial_\rho U_A,\\
   % \Gamma^v_{\rho\rho}&=0,\\
  %  \Gamma^v_{\rho A}&=0,\\
    \Gamma^v_{AB}&=-\frac{1}{2}\partial_\rho g_{AB},\\
    \Gamma^\rho_{vv}&=\frac{1}{2}F\partial_\rho \left(F-g_{AB}U^AU^B\right)-\frac{1}{2}\partial_v\left(F-g_{AB}U^AU^B\right)-U^A\partial_r U_A-\frac{1}{2}U^A\partial_A \left(F-g_{CD}U^CU^D\right),\\
    \Gamma^\rho_{v\rho}&=-\frac{1}{2}\partial_\rho F+\frac{1}{2}U_A\partial_\rho U^A,\\
    \Gamma^\rho_{vA}&=-\frac{1}{2}\partial_A\left(F-g_{BC}U^BU^C\right)-\frac{1}{2}F \partial_\rho U_A-\frac{1}{2}\partial_v g_{AB} U^B-\frac{1}{2}(\nabla_AU_B-\nabla_BU_A)U^B,\\
  %  \Gamma^\rho_{\rho\rho}&=0,\\
    \Gamma^\rho_{\rho A}&=-\frac{1}{2}\partial_r g_{AB}U^B+\frac{1}{2}\partial_r U_A=\frac{1}{2}g_{AB}\partial_\rho U^B,\\
    \Gamma^\rho_{AB}&=-\frac{1}{2}F\partial_\rho g_{AB}-\frac{1}{2}\partial_v g_{AB}+\frac{1}{2}(\nabla_AU_B+\nabla_BU_A),\\
    \Gamma^A_{vv}&=-\frac{1}{2}U^A\partial_\rho\left(F-g_{CD}U^CU^D\right)+\frac{1}{2}\nabla^A\left(F-g_{CD}U^CU^D\right)+g^{AB}\partial_v U_B,\\
    \Gamma^A_{v\rho}&=\frac{1}{2}g^{AB}\partial_\rho U_B,\\
    \Gamma^A_{vB}&=\frac{1}{2}U^A\partial_\rho U_B+\frac{1}{2}g^{AC}\partial_v g_{BC}+\frac{1}{2}g^{AC}(\nabla_BU_C-\nabla_CU_B),\\
  %  \Gamma^A_{\rho\rho}&=0,\\
    \Gamma^A_{\rho B}&=\frac{1}{2}g^{AC}\partial_\rho g_{BC},\\
    \Gamma^A_{BC}&=\frac{1}{2}U^A\partial_\rho g_{BC}+{}^{(2)}\Gamma^A_{BC}.
\end{align} In these expressions, indices $A, B, \cdots$ are raised and lowered by the metrics $g^{AB}$ and $g_{AB}$, respectively. The symbol ${}^{(2)}\Gamma^A_{BC}$ denotes the Christoffel symbol associated with the metric $g_{AB}$. Analogous to equation \eqref{csnull}, the Pontryagin term can be reduced to a Chern–Simons term on the event horizon
\be 
\text{CS}_{\text{horizon}}=\lim_{\rho\to 0,\ v\ \text{finite}}\int_{\mathcal H_\rho} d^3y \sqrt{h} n_\mu \epsilon^{\mu\nu\rho\sigma}\left(\Gamma^\alpha_{\nu\beta}\partial_\rho\Gamma^\beta_{\sigma\alpha}+\frac{2}{3}\Gamma^\alpha_{\nu\beta}\Gamma^\beta_{\rho\gamma}\Gamma^\gamma_{\sigma\alpha}\right)
\ee where $\mathcal{H}_\rho$ is the hypersurface of constant $\rho$, with normal covector $n_\mu = F^{-1/2} \delta^\rho_\mu$. The determinant of the induced metric in the coordinates $y=(v,x^A)$ is denoted by $h$. After a lengthy calculation, we find %\footnote{The quadratic terms of $\varphi$ has been canceled via integration by parts.}
\be
\text{CS}_{\text{horizon}}=\int dv d^2x \sqrt{\gamma}\mathcal{O},\label{bdyterm}
\ee 
where the operator $\mathcal{O}$ on the horizon is 
\be 
\mathcal O=-\frac{1}{2}\epsilon^{AB}\overline U_A\dot{\overline U}_B-\kappa\epsilon^{AB}\partial_A\overline U_B-\epsilon^{AB}\overline{U}_A\partial_B\kappa-\dot\varphi \epsilon^{AB}\nabla_B\overline U_A-\epsilon^{AB}\overline U_B\nabla_A\dot\varphi+2\epsilon^{AB}\partial_A\varphi\partial_B\dot\varphi,\label{operaterO}
\ee 
where $\dot{\overline U}_A=\partial_v{\overline U}_A$. After integration by parts, the Chern–Simons term on the horizon becomes
\be 
\text{CS}_{\text{horizon}}=\int dv d^2x \sqrt{\gamma}[-\frac{1}{2}\epsilon^{AB}\overline U_A\dot{\overline U}_B+(\dot\varphi-\kappa)\epsilon^{AB}(\partial_A\overline U_B-\partial_B\overline U_A)].\label{cshorizon}
\ee 

Interestingly, the above term \eqref{cshorizon} vanishes when the twist 1-form $\overline U_A=0$. The first term in the square bracket is a quadratic function of the twist field whose form is rather similar to the electromagnetic helicity flux \cite{Liu:2024rvz}. The second term is basically products of $(\dot\varphi-\kappa)$ and the unique 2-form field on the transverse direction constructed by the twist field. One can simplify the expression further by imposing the vacuum Einstein equation. At the leading order, the Einstein equation is 
\begin{align}
    0&= \dot\varphi \kappa- \left(\dot\varphi^2+\ddot\varphi\right),\\
    0&=\partial_A\kappa+\frac{1}{2}\dot{\overline U}_A+\overline U_A\dot\varphi+\partial_A\dot{\varphi},\label{Damour1}\\
    0&=-F_2+\frac{1}{2}\nabla_C\overline U^C+\frac{1}{2}\overline U_A\overline U^A-\frac{1}{2}\kappa \lambda^C_{\ C}+\frac{1}{4}\lambda^{AB}\dot\gamma_{AB}-\frac{1}{2}\gamma^{AB}\dot\gamma_{AB},\\
    0&=\lambda_{AB}\lambda^{AB}-4\gamma^{AB}k_{AB},\\
    0&={}^{(2)}\overline U_A-\frac{1}{2}\overline U^C\lambda_{AC}+\frac{1}{2}\nabla_B\lambda^B_{\ A}-\frac{1}{2}\nabla_A\lambda^C_{\ C}+\frac{1}{4}\lambda^C_{\ C}\overline U_A,\\
    0&=-\dot\lambda_{AB}-\kappa\lambda_{AB}+\frac{1}{2}(\nabla_A\overline U_B+\nabla_B\overline U_A)+\overline R_{AB}-\frac{1}{4}\gamma^{CD}\dot\gamma_{CD}\lambda_{AB}-\frac{1}{4}\lambda^C_{\ C}\dot\gamma_{AB}\nn\\&\hspace{0.5cm}-\frac{1}{2}\lambda^C_{\ A}\dot\gamma_{BC}-\frac{1}{2}\lambda^C_{\ B}\dot\gamma_{AC}-\frac{1}{2}\overline U_A\overline U_B.
\end{align} We notice that the first equation is the null Raychaudhuri equation \cite{1955PhRv...98.1123R} and the second  is the Damour equation \cite{1982mgm..conf..587D}. There is a trivial solution for the null Raychaudhuri equation 
\be 
\dot\varphi=0\label{dotph}
\ee and the Chern-Simons term becomes
\bea  
\text{CS}_{\text{horizon}}&=&\int dv d^2x \sqrt{\gamma}[-\frac{1}{2}\epsilon^{AB}\overline U_A\dot{\overline U}_B-\kappa\epsilon^{AB}(\partial_A\overline U_B-\partial_B\overline U_A)]\label{csexpansion0}
\eea 
Note that the result \eqref{cshorizon} only depends on the near-horizon geometry and is independent of GR while the above quantity is valid under the condition \eqref{dotph}. %Therefore, we will leave it as \eqref{cshorizon} without imposing the equation of GR. %Readers can find more discussions in Appendix \ref{simp} on the simplification of the Chern-Simons term \eqref{cshorizon} using Einstein equation.

As a consistency check, we will also use the tetrad formalism to evaluate the same boundary term.  We choose the vielbein for the near-horizon geometry 
\bs
\begin{align}
     \theta^0=&\frac{1}{2}(-1-F)dv+d\rho,\\ \theta^1=&\frac{1}{2}(-1+F)dv-d\rho,\\ \theta^{\hat{A}}=&E^{\hat{A}}_Adx^A+U^{\hat{A}}dv,
\end{align}\es
and the leading terms of the corresponding spin connection are
\bs 
 \begin{align}
        \omega^{01}=&{\kappa dv-\frac{1}{2}\overline U_A dx^A+\mathcal{O}(\rho)},\\
        \omega^{0\hat{A}}=&{\frac{1}{4}\overline U^{\hat A} dv+\frac{1}{2}\overline U^{\hat A}d\rho+(-\dot\varphi \delta_A^{\hat A}+\frac{1}{4}\lambda_A^{\hat A})dx^A+\mathcal{O}(\rho)},\\
        \omega^{1\hat{A}}=&{\frac{1}{4}\overline U^{\hat A}dv-\frac{1}{2}\overline U^{\hat A}d\rho +(\dot\varphi \delta_A^{\hat A}+\frac{1}{4}\lambda_A^{\hat A})dx^A+\mathcal{O}(\rho)},\\
        \omega^{\hat{A}\hat{B}}=&{-\epsilon^{\hat A\hat B}\epsilon_{AB}\partial^A\varphi dx^B+\mathcal{O}(\rho)}\label{omegaAB}.
    \end{align}\es
Fields with Lorentz indices $\hat{A}$ are the projections of the spacetime fields onto the local Lorentz frame by means of the tetrad. It is straightforward to check that the formula \begin{align}
    \text{CS}_{\text{horizon}}=&\lim_{r\to\infty,\ v \ \text{finite}}\int_{\mathcal H_r}
 \left(\omega^{ab}\wedge d\omega_{ab}-\frac{2}{3}\omega^a_{\ b}\wedge \omega^b_{\ c}\wedge \omega^c_{\ a}\right)\end{align} leads to the same result as \eqref{cshorizon}. 
 %Then the  equivalent form of the boundary Chern-Simons term is
%\begin{align}
 %   \text{CS}_{\text{horizon}}=&\lim_{r\to\infty,\ v \ \text{finite}}\int_{\mathcal H_r}
% \left(\omega^{ab}\wedge d\omega_{ab}-\frac{2}{3}\omega^a_{\ b}\wedge \omega^b_{\ c}\wedge \omega^c_{\ a}\right)\nn\\
% =&-\int dv\wedge dx^A\wedge dx^B(-\frac{1}{2}\overline U_A\dot{\overline U}_B-\kappa\partial_A\overline U_B-\overline U_A\partial_B\kappa-\dot{\varphi}\partial_B\overline U_A-\overline U_B\partial_A\dot{\varphi})\label{spinconnection},
%\end{align}
%which is the same as \eqref{cshorizon} after integration by parts up to a sign. %$\om_{\hat{A}\hat{B}}\partial_v\om^{\hat{A}\hat{B}}$, which is eliminated in advance through antisymmetry in \eqref{omegaAB}.

 \paragraph{Comparison to the fluid helicity.}
We now explore the physical implications of the boundary term \eqref{cshorizon}. Interpreting this quantity from the perspective of Carrollian hydrodynamics is particularly interesting. We would like to use the Papapetrou–Randers (PR) coordinates on three-dimensional spacetime $\overline M$%\cite{}
\begin{align}
    ds^2=-c^2(\alpha dv-b_Adx^A)^2+\gamma_{AB}dx^Adx^B \label{line},
\end{align} where $c$ is the velocity of light. In the limit of $c\to 0$, it is exactly the metric of a Carrollian manifold. The field $b_A$ becomes the Ehresmann
connection and kernel of the Carrollian manifold is $\bm\chi=\frac{1}{\alpha}\partial_v$ whose dual form is 
\be 
\bm\sigma=-\alpha dv+b_Adx^A. \label{matchu2}
\ee As we will see, this dual form is related to the velocity of the Carrollian fluid $\bm u$ up to a factor. From an intrinsic perspective, the authors of \cite{Ciambelli:2018xat} argued that the fluid velocity may be parameterized as 
\be 
u_0=-\frac{c^2 \alpha}{\sqrt{1-c^2\bm\beta^2}},\quad  u_A=\frac{c^2(b_A+\beta_A)}{\sqrt{1-c^2\bm\beta^2}}\label{parau}
\ee where $\beta_A$ is  a Carroll parameter whose dimension is the inverse velocity \cite{Duval:2014uoa} with $\bm\beta^2=\beta_A\beta_B\gamma^{AB}$. One can check the normalization $\bm u^2=c^2$ and  the velocity of the fluid vanishes in the limit $c\to 0$. The expression \eqref{parau} matches the dual form \eqref{matchu2} for $\beta_A=0$. The factor $c$ should be sent to zero for strictly Carrollian fluid. Therefore, we may renormalize the velocity to the form 
\be 
\bm u_{\text{r}}=-\alpha dv+(b_A+\beta_A)dx^A\label{vform}=\bm\sigma
\ee 
for strictly Carrollian fluid. In both relativistic hydrodynamics and its non-relativistic limit, fluid helicity \cite{1984JFM...147..133B} is a topological number characterizing the linkage and twist of vorticity in the fluid. This concept is closely related to earlier results on magnetic helicity in magnetohydrodynamics \cite{elsasser1956hydromagnetic, woltjer1958theorem}. Given a vector field $\bm X$ and its vorticity $\nabla\times\bm X$ in three-dimensional space, the pseudoscalar of the form \cite{1969JFM....35..117M}
\be 
\int d^3\bm x \bm X\cdot(\nabla\times\bm X) \label{top}
\ee characterizes the topological property of the field lines. By setting $\bm{X} = \bm{a}$, where $\bm{a}$ is the magnetic vector potential, the associated magnetic field becomes $\bm{b} = \nabla \times \bm{a}$. Consequently, the topological number \eqref{top} is identically the magnetic helicity
\be 
\mathcal H_{m}=\int d^3\bm x\ \bm a\cdot\bm b.
\ee In fluid dynamics, letting $\bm{X} = \bm{v}$ (with $\bm{v}$ denoting the fluid velocity), one identifies the topological number \eqref{top} with the fluid helicity
\be 
\mathcal H_f=\int d^3\bm x\ \bm v\times (\nabla\times\bm v).
\ee 
Recently, the topological number \eqref{top} has been discussed using Chern-Simons term \cite{Liu:2024rvz} 
\be 
\text{CS}[\bm a]=\int_{\overline M} \bm a\wedge d\bm a \label{cha}
\ee where $\overline{M}$ is a three-dimensional manifold. When $\overline M$ is a constant time slice, the Chern-Simons term reduces  to the magnetic helicity. When $\overline M$ is future null infinity, the Chern-Simons term induces the so-called EM helicity flux \cite{Liu:2023qtr}. Now we extend the concept to Carrollian fluid. Note that we have a velocity 1-form \eqref{vform} in the three dimensional Carrollian manifold $\overline M$ ($c\to 0$ limit). This manifold has a natural Chern-Simons term 
\bea 
\mathcal H_{C}=\lim_{c\to 0}\frac{1}{c^4}\int_{\overline M}\bm u\wedge d\bm u\label{HC}
\eea where the subscript $C$ is a shorthand for Carroll. We will refer to this term as the Carrollian fluid helicity, which takes the form of
\begin{align}
\mathcal H_C=& -\int dvd^2x\sqrt{\gamma}\epsilon^{AB}(\alpha\partial_Ab_B+b_A\dot{b}_B+b_A\partial_B\alpha+\alpha\partial_A\beta_B+\beta_A\dot{b}_B+b_A\dot{\beta}_B+\beta_A\dot{\beta}_B+\beta_A\partial_B\alpha),\label{Khelicity}
\end{align}
where $\dot{b}_B=\partial_vb_B$  and $\dot{\beta}_B=\partial_v\beta_B$.
In general, one may multiply the velocity $\bm u$ by a function $f$, 
\be 
\bm u'=f \bm u,
\ee then the corresponding vorticity becomes 
\be 
\bm \omega'=d\bm u'=f\bm \omega+df\wedge\bm u.
\ee 
Using the function $f$, we define a modified Carrollian fluid helicity
\be 
\mathcal{H}_C'=\lim_{c\to 0}\frac{1}{c^4}\int_{\overline M}\bm u'\wedge d\bm u'=\lim_{c\to 0}\frac{1}{c^4}\int_{\overline M}f^2 \bm u\wedge d\bm u.\label{modHC}
\ee Unlike the Carrollian fluid helicity \eqref{HC}, this variant contains an additional function $f^2$ within the integral. Nevertheless, the $df$ term in the modified vorticity makes no contribution to the modified Carrollian fluid helicity.  

Introducing the function $f$ is permissible since vorticity can be defined in multiple ways within relativistic hydrodynamics based on the context. According to \cite{Becattini:2015ska,Deng:2016gyh}, the standard vorticity associated with the velocity $\bm u$ is called the kinematical vorticity. By choosing $f$ to be the temperature $T$, one can define the so-called $T$-vorticity and the relevant helicity has also been constructed\cite{Huang:2020dtn}. There are other different definitions such as thermal vorticity $f=\frac{1}{T}$ and enthalpy vorticity $f=w$, with $w$ the specific enthalpy. One of the original intentions of these definitions is to satisfy relativistic Helmholtz-Kelvin theorem \cite{10.1093/acprof:oso/9780198528906.001.0001} which states that the vortex lines move with the fluid as if they are frozen in the fluid.  As an illustration, we consider a relativistic ideal neutral fluid that satisfies the Euler equation \begin{align}
    (\varepsilon+P)\frac{d}{d\tau}u^\mu=d^\mu P\quad\Rightarrow\quad\frac{d}{d\tau}(Tu^\mu)=\partial^\mu T\label{Euler}
\end{align} 
\iffalse 
{\zxh
\begin{align}
    (\varepsilon+P)\frac{d}{d\tau}u^\mu=d^\mu P\quad&\Rightarrow \varepsilon\frac{d}{d\tau} u^\mu+\frac{d}{d\tau}(Pu^\mu)=\partial^\mu P\\
    &\Rightarrow s\frac{d}{d\tau}(Tu^\mu)=s\partial^\mu T \quad\text{for} \quad dP=sdT,sT=\varepsilon+P
\end{align}
}\fi
where $\varepsilon$ and $P$ are the mass density and pressure of the fluid, respectively. The four velocity $u_\mu$ can be parameterized as $\gamma(1,v^i)$ and $\frac{d}{d\tau}=u^\mu\partial_\mu$ is the proper time or comoving time derivative and $d_\mu=\partial_\mu-u_\mu\frac{d}{d\tau}$. The equation \eqref{Euler} leads to the relativistic Helmholtz-Kelvin theorem
\begin{align}
    \frac{d}{d\tau}\oint_{\Gamma} T u_\mu dx^\mu=\oint_{\Gamma}\partial_\mu Tdx^\mu=0
\end{align} which means that the circulation integral around a closed loop $\Gamma$ of an effective velocity $u_\mu'=Tu_\mu$ is conserved. Hence one may choose $f=T$ for a relativistic  ideal neutral fluid to construct the T-vorticity. 
 
 In the non-relativistic limit $c\xrightarrow{}\infty$ with coordinates $(t,x^i)$, we have a definition of vorticity for the fluid with the 3-velocity $\bm v$
\begin{align}
    \bm\omega=\nabla\times (f\bm v), \quad f=1.
\end{align}
For ideal incompressible fluid $(\nabla\cdot\bm v=0)$, by using the evolution equation, we have the non-relativistic Helmholtz-Kelvin theorem 
\begin{align}
    \frac{d}{dt}\oint_\Gamma \bm v \cdot d\bm x=0,
\end{align}where $\Gamma$ is a closed contour. %The
%circulation conservation means that the vortex lines are co-moving with the fluid, as if they are frozen in the fluid \cite{Deng:2016gyh}. 

We now turn to a Carrollian fluid whose dynamics are captured by the near-horizon geometry.
\iffalse 
We impose the condition \eqref{dotph} and the Damour equation becomes
\be 
\frac{1}{2}\dot{\overline U}_A+\partial_A\kappa=0.
\ee \fi According to the identification \cite{Donnay:2019jiz}, we can determine the functions in the Carrollian fluid velocity by the near-horizon quantities
\bea 
\alpha=\sqrt{2\kappa},\quad b_A=\frac{\overline U_A}{\sqrt{2\kappa}},\quad c=\sqrt{\rho}\label{alphabA}
\eea up to a Carroll parameter $\beta_A$. The modified Carrollian fluid helicity contains an additional unfixed function $f$. We find that the modified Carrollian fluid helicity matches the boundary term \eqref{bdyterm}-\eqref{operaterO}, provided that
\be 
f=\sqrt{\kappa},\quad \beta_A=-\sqrt{\frac{2}{\kappa}}\partial_A\varphi.\label{fbetaA}
\ee 
\paragraph{Remarks.} 
\begin{enumerate}
    \item The function $f = \sqrt{\kappa}$ is particularly interesting. For a stationary black hole, the surface gravity $\kappa$ is proportional to the Hawking temperature. Consequently, this case differs from the definition of $T$-vorticity with $f = T$. To understand the relation $f = \sqrt{\kappa}$, we simplify the discussion by setting $\beta_A = 0$. Considering the non-expanding solution \eqref{dotph}, the Damour equation reduces to \be 
\frac{1}{2}\dot{\overline U}_A+\partial_A\kappa=0.\label{Damoursim}
\ee We may construct a circulation integral $ 
\oint_{\Gamma}f\bm u
$ around a contour $\Gamma$ whose time-derivative is 
\be 
\frac{d}{dv}\oint_{\Gamma}f\bm u=\oint_{\Gamma}\partial_v(f b_A)dx^A.
\ee An ultra-relativitic extension of the Helmholtz-Kelvin theorem would require
\be 
\partial_v \left(f\frac{\overline {U}_A}{\sqrt{2\kappa}}\right)=\partial_A(\cdots)\label{HKe}
\ee where $\cdots$ are smooth functions on the horizon. By setting $f=f_0\sqrt{\kappa}$ where $f_0$ is a non-vanishing constant and utilizing the Damour equation \eqref{Damoursim}, the conservation of the circulation integral \eqref{HKe} is established. We may choose the constant $f_0=1$ \footnote{This constant can always be absorbed into the normalization of Carrollian fluid helicity.} and thus the fluid velocity 1-form $\bm u'=\sqrt{\kappa}\bm u$ may be used to construct the modified Carrollian fluid helicity. Note that the corresponding modified vorticity is \be \bm\omega'=d (f\bm u)=d(-\sqrt{2}\kappa dv+\frac{\overline U_A}{\sqrt{2}}dx^A)\ee whose spatial component $\omega'_{AB}$ matches the one in \cite{Redondo-Yuste:2022czg} up to a constant number. However, in our case, there are also non-vanishing components  $\omega'_{vA}$.

There is a direct relation between the velocity 1-form $\bm u'$ and the spin connection\footnote{We thank the anonymous referee for this valuable suggestion.} for constant $\varphi$. In this case, the spin connection that contributes to CS$_{\text{horizon}}$ is
    \be 
    \omega^{01}=\kappa dv-\frac{1}{2}\overline U_A dx^A.
    \ee As a consequence, the boundary term 
    \bea 
    \text{CS}_{\text{horizon}}=\lim_{\rho\to0,\ v \ \text{finite}}2\int \omega^{01}\wedge d\omega_{01}\label{CS01}
    \eea has the same form as \eqref{modHC}. More precisely, 
    \be 
    \lim_{c\to 0}\frac{1}{c^2}\bm u'=f\left(-\sqrt{2\kappa}dv+\frac{\overline U_A}{\sqrt{2\kappa}}dx^A\right).
    \ee The component $\omega_{01}$ can be mapped to the velocity 1-form 
    \be 
   \lim_{\rho\to 0,\ v\ \text{finite}} \omega_{01}=\lim_{c\to 0}\frac{1}{c^2}\bm u'
    \ee via the identification 
    \be 
    f=\sqrt{\kappa}.
    \ee 
    Note that the expression \eqref{CS01} does not hold for more general $\varphi$. In that case, other components of the spin connection also contribute to $\text{CS}_{\text{horizon}}$, and $\bm{u}'$ is modified by the Carroll parameter $\beta_A$ accordingly. %the cannot be mapped to the velocity 1-form.

    \item The identification $\beta_A = -\sqrt{\frac{2}{\kappa}} \partial_A \varphi$ has not been mentioned in the literature. For constant $\varphi$, the Carroll parameter yields $\beta_A = 0$. 
    However, since $\varphi$ is generally angle-dependent, $\beta_A$ is typically non-vanishing. Hence, $\beta_A$ characterizes the inhomogeneous properties of the background manifold, as $\varphi$ determines the metric of the Carrollian structure. The inhomogeneity of the Carrollian manifold prevents the Carrollian fluid velocity from being uniform, making it impossible to set $\beta_A = 0$.
\end{enumerate}

\section{Conclusion and discussion}\label{conc}
%The role of torsion in physics \cite{2002PhR...357..113S}. Gravitational theory with torsion \cite{RevModPhys.48.393}.
%References on gravitational anomaly \cite{Kimura:1969iwz,Delbourgo:1972xb,Alvarez-Gaume:1983ihn}. Nieh-Yan term as a difference of Pontryagin terms \cite{PhysRevD.55.7580}. Debate on torsion contribution to chiral anomaly \cite{Rasulian:2023gtb}
In this work, we reduce two topological invariants, the Pontryagin and Nieh-Yan terms, to future null infinity and near-horizon geometry. Combined with our previous work in electromagnetics, results are summarized in Table \ref{summary}.
\begin{table}
\begin{center}
\renewcommand\arraystretch{1.5}
 \begin{tabular}{|c|c||c|c|}\hline
Topological term&Theory &Hypersurface& Observable\\\hline\hline
Chern number&EM&Constant time slice&Magnetic helicity\\\hline
Chern number&EM&Future null infinity&EM helicity flux \\\hline
Pontryagin term & GR & Future null infinity&$\times$ \\\hline
Pontryagin term &GR&Horizon & Carrollian fluid helicity \\\hline
Nieh-Yan term &TEGR&Future null infinity&Gravitational helicity flux \\\hline
Nieh-Yan term &TEGR&Horizon&$\times$\\\hline
\end{tabular}
\caption{\centering{Reduction of topological invariants and observables}}\label{summary}
\end{center}
\end{table} 
\begin{enumerate}
    \item \textbf{Chern number.} The first two lines are mainly discussed in our work \cite{Liu:2024rvz} and we have extended it to higher dimensions\footnote{See also a brief discussion in \cite{Long:2024yvj}.}. The Chern number can  be reduced to a constant time slice and future null infinity, then the boundary quantity would be magnetic helicity and EM helicity flux, respectively. We also notice the possibility to reduce the Chern number to spatial infinity (see the last term on the right hand side of equation (2.52) in \cite{Heng:2025kmr}) and more general timelike hypersurfaces. However, the physical interpretation is unclear so far and we don't include those cases in the table.
    \item \textbf{Pontryagin term.} One of the main motivations is to reduce the Pontryagin term to future null infinity and reproduce the gravitational helicity flux. Unfortunately, the standard fall-off condition for gravitational waves leads to a vanishing result. Therefore, we add a symbol $\times$ in the third line. Interestingly, one can reduce Pontryagin term to a dynamical horizon and find a non-trivial observable \eqref{cshorizon}. The observable can be described by the surface gravity, expansion and twist field from the near-horizon geometry which is closely related to the Carrollian fluid. We interpret this term as a modified  Carrollian fluid helicity 
    \begin{align}
\mathcal H'_C=& -\int dvd^2x\sqrt{\gamma}f^2\epsilon^{AB}(\alpha\partial_Ab_B+b_A\dot{b}_B+b_A\partial_B\alpha+\alpha\partial_A\beta_B+\beta_A\dot{b}_B+b_A\dot{\beta}_B+\beta_A\dot{\beta}_B+\beta_A\partial_B\alpha),\label{mKhelicity}
\end{align}
    with the identifications \eqref{alphabA} and \eqref{fbetaA}. It would be wonderful to understand the identification \eqref{fbetaA} from other perspectives.
    \item \textbf{Nieh-Yan term.} Interestingly, it is the Nieh-Yan term in TEGR that could induce the gravitational helicity flux. This has been summarized in the fifth line of the table. The result supports the point of view that Nieh-Yan term could be useful to find physical observables since gravitational helicity flux density is an observable in GR. Unfortunately, this result is obtained in TEGR and it should be interpreted more carefully. Maybe one should reconsider the role of torsion in physics? In the last decades,  modified gravity in terms of torsion, such as $f(T)$ gravity\footnote{$f(T)$ gravity is a torsion-based modified gravity that generalizes the action of TEGR by replacing the torsion scalar $T$ 
with an arbitrary function $f(T)$,
which is similar in spirit to the generalization of the Ricci
scalar $R$ in the Einstein-Hilbert action to a function $f(R)$\cite{Capozziello:2011et}. Note that in TEGR, as well as $f(T)$ gravity, the torsion entirely characterizes the non-trivial geometry of spacetime. In contrast, the torsion in EC gravity is an additional non-dynamical field to curvature. }  has been applied to cosmology to address problems such as dark matter and dark energy \cite{Bengochea:2008gz,Cai:2015emx}. By reducing the Nieh-Yan term to a general horizon in TEGR, we don't find non-trivial results. Therefore, we add a $\times$ in the last line\footnote{We adopt the conformal gauge (3.8) for the metric $\gamma_{AB}$, following references [55, 97, 98]. In this gauge, the boundary term $H=\Lambda^2\int_{\overline M}\theta^a\wedge T_a$ vanishes in the pure tetrad formulation $\Omega_{\mu}^{\ ab}=0$. In general, however, a non-vanishing result may be obtained, depending on the choices of gauge for both the metric and the spin connection. This issue is not yet well understood and is left for future work.}. In gravitational physics, null infinity and horizons correspond to infrared (IR) and ultraviolet (UV) physics, respectively. The results indicate that the Nieh-Yan term is important for IR physics. A thermal Nieh-Yan term has also been proposed in condensed matter physics \cite{Nissinen:2019mkw,Nissinen:2019wmh}, where the cutoff $\Lambda$ is replaced by the well-defined IR scale temperature $T$ through the relation:
\begin{align}
\Lambda^2 = \frac{1}{12}T^2.
\end{align} On the other hand, the Pontryagin term is important for UV physics but irrelevant to IR physics, as it yields a vanishing result at null infinity and a non-trivial result on the horizon. 
    \item \textbf{Euler number.} In four dimensions, the Euler number is exactly the Gauss-Bonnet term. We don't pay much attention to this topological invariant since it is not directly related to chiral anomaly \cite{Bilal:2008qx,Harvey:2005it}. Therefore, we don't expect that it could be used to deduce the physical observables that relate to helicity.  However, we notice that Euler number is associated with Weyl anomaly and the a-theorem \cite{Duff:1977ay,Duff:1993wm,1988PhLB..215..749C,Grassi:2011hq}. It would be interesting to explore whether its reduction on certain kinds of hypersurfaces could lead to new observables.
\end{enumerate}

%The role of torsion on teleparallel gravity .
%\cite{Cartan1979ElieCA}{\zxh It has been shown that recently detected acceleration of the universe can be understood by dark torsion, with no need of dark energy\cite{Bengochea:2008gz}. }
\textbf{Note added.} In  \cite{PhysRevD.104.065012}, the authors found another non-vanishing  observable at future null infinity to test chiral aspects of gravitational waves \cite{PhysRevLett.134.031402}
\be 
\widetilde{\text{CS}}_+=\int du d\Omega \ddot{C}_{AB}\epsilon^{BC}\dot C_{C}^{\ A}\label{ch}
\ee from the Pontryagin term.
The above result is only one part of  the boundary  term coming either from 
\be 
\int du \wedge dx^A\wedge dx^B \omega_{A\hspace{3pt}\hat{ B}}^{\ 0}\wedge \dot{\omega}_{B\hspace{3pt}0}^{\ \hat B}
\ee or 
\be 
\int du\wedge dx^A\wedge dx^B \omega_{A\hspace{3pt}\hat{ B}}^{\ 1}\wedge \dot{\omega}_{B\hspace{3pt}1}^{\ \hat B}.
\ee However, the summation of these two terms cancels, as has been checked using alternative form \eqref{csnull}.
In our work, it is the Nieh-Yan term that leads  to a slightly different parity-odd quantity (gravitational helicity flux) \cite{Seraj:2022qyt,Liu:2023gwa}
\be 
H=\frac{1}{32\pi G}\int du d\Omega \epsilon^{BC}C_{AB}\dot C_C^{\ A}
\ee whose density can also test the chiral aspects of gravitational waves.

 \vspace{3pt}
{\bf Acknowledgments.} 
The work of J.L. was supported by NSFC Grant No. 12005069.
%\appendix
%\section{Chern-Simons term and Einstein equation}\label{simp}
\iffalse 
The starting point is the Damour's equation 
\be 
\frac{1}{2}\dot{\overline U}_A=-\overline U_A\dot\varphi-\partial_A(\kappa+\dot\varphi).
\ee It follows that 
\be 
\frac{1}{2}\int dv d^2x\sqrt{\gamma} (\kappa+\dot\varphi)\epsilon^{AB}(\partial_A\overline U_B-\partial_B\overline U_A)=\int dv d^2x \sqrt{\gamma}\epsilon^{AB} \overline U_B\partial_A(\dot\varphi+\kappa)=\frac{1}{2}\int dv d^2x \sqrt{\gamma}\epsilon^{AB} \overline U_A\dot{\overline U}_B.
\ee  Substituting it into \eqref{cshorizon} and eliminating the quadratic term of the twist field, we find 
\bea 
\text{CS}_{\text{horizon}}=\frac{1}{2}\int dv d^2x \sqrt{\gamma}[(3\dot\varphi-\kappa)\epsilon^{AB}(\partial_A\overline U_B-\partial_B\overline U_A)].\label{a3}
\eea One can also eliminate the term involving $\dot\varphi$ and then 
\be 
\text{CS}_{\text{horizon}}=\int dv d^2x \sqrt{\gamma}[-\frac{3}{2}\epsilon^{AB}\overline U_A\dot{\overline U}_B-2\kappa \epsilon^{AB}(\partial_A\overline U_B-\partial_B\overline U_A)].\label{a4}
\ee The two expressions \eqref{a3} and \eqref{a4} are equivalent by imposing the Damour's equation.\fi
\bibliography{refs}
\end{document}